\documentclass[%
 reprint,
showpacs,
 amsmath,amssymb,
 aps,
]{revtex4-1}

\usepackage[pdftex]{graphicx}
\usepackage{dcolumn}
\usepackage{bm}
\usepackage{color}
\usepackage{slashbox}

\begin{document}


\title{Roles of $^7$Be$(n,p)^7$Li resonances in big bang nucleosynthesis with time-dependent quark mass and Li reduction by a heavy quark mass}

\author{Kanji Mori}
\affiliation{%
Department of Astronomy, Graduate School of Science, The University of Tokyo, 7-3-1 Hongo, Bunkyo-ku,
Tokyo, 113-0033 Japan}%
\affiliation{%
National Astronomical Observatory of Japan, 2-21-1 Osawa, Mitaka, Tokyo,
181-8588 Japan}%

\author{Motohiko Kusakabe}
\affiliation{
School of Physics, Beihang University, 37 Xueyuan Road, Haidian-qu, Beijing
100083, China
}%

\date{\today}

\begin{abstract}
Big bang nucleosynthesis (BBN) has been used as a probe of beyond-standard physics in the early Universe, which includes a time-dependent quark mass $m_q$. We investigate effects of a quark mass variation $\delta m_q$ on the cross sections of the  $^7$Be$(n,p)^7$Li reaction and primordial light element abundances taking into account roles of  $^8$Be resonances in the reaction during BBN. This resonant reaction has not been investigated although behaviors of low-lying resonances are not trivial. It is found that a resonance at the resonance energy $E_{\rm r}$=0.33 MeV enhances the reaction rate and lowers the $^7$Li abundance significantly when the quark mass variation is negative. Based upon up-to-date observational limits on primordial abundances of D, $^4$He and Li, the quark mass variation in the BBN epoch are derived. In a model in which the resonance energies of the reactions $^3$He$(d, p)^4$He and $^3$H$(d, n)^4$He are insensitive to the quark mass, we find that the Li abundance can be consistent with observations for $\delta m_q/m_q =($4--8$)\times 10^{-3}$.
\end{abstract}

\pacs{26.35.+c, 21.10.Dr, 98.80.Cq, 98.80.-k}

\maketitle


The Standard Model of particle physics assumes that fundamental constants are time-independent over the cosmic history. However, possibility of time-dependent constants including bare coupling constants have been pursued for a long time \cite{dirac,uzan}. 

The standard big bang nucleosynthesis (BBN) model \cite{kolb,cyburt,pitrou} is characterized by a single cosmological parameter, the baryon-to-photon ratio. This parameter is precisely measured by cosmic microwave background observations \cite{Hinshaw:2012aka,planck}, and now BBN is a useful probe of beyond-standard physics in the early Universe. In addition, it has been pointed out that the observed $^7$Li abundance of metal-poor stars \cite{spite} is significantly lower than the standard BBN (SBBN) prediction \cite{Ryan:1999vr,Melendez:2004ni,Asplund:2005yt,Bonifacio:2006au,shi2007,Aoki:2009ce,Sbordone:2010zi}. The solution to this ``lithium problem'' may come from systematic errors in inferring the primordial abundance from astronomical observations or lack of nuclear cross section data \cite{Cyburt:2009cf,Chakraborty:2010zj}. However, a reasonable solution has not been verified neither from astronomy nor from nuclear physics. Hence it is necessary to explore non-standard BBN to solve this problem.

BBN with a time-dependent quark mass has been studied by several authors \cite{berengut,cheoun, bedaque}. A quark mass variation affects nuclear binding energies and hence resonance energies. Among main reactions in the BBN reaction network, $^3$He$(d, p)^4$He, $^3$H$(d, n)^4$He, and $^7$Be$(n, p)^7$Li are the only reactions whose cross sections are governed by resonances \cite{descouvemont}. The resonances in the reactions $^3$He$(d, p)^4$He and $^3$H$(d, n)^4$He have been treated \cite{berengut,cheoun, bedaque}. The $^7$Be$(n, p)^7$Li reaction is by far the strongest destruction reaction for $^7$Be nuclei in SBBN \cite{kawano}. Therefore, the reaction cross section has been measured by many nuclear experiments \cite{exp1,exp2,exp3,exp4,exp5,exp6}. However, resonances in the reaction $^7$Be$(n, p)^7$Li have been ignored in the model of varying quark mass since the reaction is dominated by a broad resonance located around the separation threshold energy of the entrance channel. Nevertheless, a behavior of the near-threshold broad resonance at $E_\mathrm{r}=2.67$ keV should be treated carefully and another resonance at $E_\mathrm{r}=0.33$ MeV can play a role when the quark mass is changed, as shown in this letter.

Recently, new observations and reanalysis of astronomical data largely reduced uncertainties in primordial abundances \cite{sbordone,cooke,izotov,aver}. In this letter, we show the most stringent constraint on quark mass in the BBN epoch using these updated observational results.

The sensitivity of the binding energy of nucleus $A$, $E_{A}$, to the quark mass variation from the present value $\delta m_q$ is parameterized in terms of a sensitivity coefficient
$K_A=\frac{\delta E_A/E_A}{\delta m_q/m_q}$,
where $m_q$ is the quark mass and $\delta E_A$ is the variation of the binding energy due to  $\delta m_q$. The $K_A$ values are adopted from Ref. \cite{flambaum} for the case with the Argonne $v_{18}$ potential used as a two-nucleon potential and the Urbana model IX used as a three-nucleon potential.

For radiative capture reactions, usually the strongest electric dipole transition dominates, and cross sections depend on the kinetic energy $E$ and the $Q$-value \cite{berengut,cheoun} as
  $\sigma(E)\propto E_\gamma^3\sim(Q+E)^3$,
where $E_\gamma$ is the energy of the emitted photon.

On the other hand, if reactions of two charged nuclei produce two charged nuclei in the final states, cross sections are proportional to the final state velocity $v\propto\sqrt{Q+E}$ and the penetration factor:
\begin{eqnarray}
\sigma(E)\propto\sqrt{Q+E}\exp\left(-\sqrt{\frac{E_\mathrm{G}}{Q+E}}\right),
  \label{eq3}
\end{eqnarray}
where $E_\mathrm{G}=2\pi^2\mu_{34} c^2(\alpha Z_3Z_4)^2$ is the Gamow energy for the exit channel with $\mu_{34}$ the reduced mass and $Z_3$ and $Z_4$ the exit-channel charges.

The $Q$ value is changed to $Q+\delta Q$ with the quark mass variation. When $E\ll Q$, which is usually realised in the BBN temperature, Eq. (\ref{eq3}) can be expanded as
\begin{eqnarray}
\frac{\sigma(E)}{\sigma_0}&\propto&\left[1+\frac{1}{2}\left(1+\sqrt{\frac{E_\mathrm{G}}{Q}}\right)\frac{\delta Q}{Q}+O\left(\left(\frac{\delta Q}{Q}\right)^2\right)\right],~~~~~
\end{eqnarray}
where $\sigma_0$ is the cross sections for the case without quark mass variation, i.e., the SBBN values.

For narrow resonances, the reaction rates with a quark mass variation are given \cite{cheoun} by
\begin{eqnarray}
N_\mathrm{A}\langle\sigma v\rangle&=&[N_\mathrm{A}\langle\sigma v\rangle]_0\left[1+\frac{1}{2}\left(1+\sqrt{\frac{E_\mathrm{G}}{Q}}\right)\frac{\delta Q}{Q}\right]\nonumber\\
&\times&\biggl\{\frac{1+[(E_0-E_\mathrm{r}^0)/(\Gamma_\mathrm{r}/2)]^2}{1+[(E_0-E_\mathrm{r})/(\Gamma_\mathrm{r}/2)]^2}\biggr\}, \label{narrow_res}
\end{eqnarray}
where $N_\mathrm{A}$ is the Avogadro number, $\langle\sigma v\rangle$ is the reaction rates per particle pair, $[N_\mathrm{A}\langle\sigma v\rangle]_0$ is the reaction rate for the case without quark mass variation, $E_0=E_\mathrm{G}^{1/3}(k_{{\rm B}}T/2)^{2/3}$ is the Gamow peak with $k_\mathrm{B}$ the Boltzmann constant and $T$ the temperature, $\Gamma_\mathrm{r}$ is the total resonance width, and $E_\mathrm{r}$ and $E^0_\mathrm{r}$ are the resonance energy for the cases with and without a quark mass variation, respectively.

For broad resonances, Eq. (\ref{narrow_res}) cannot be applied. The resonant cross sections are written by the Breit-Wigner formula \cite{iliadis}
\begin{eqnarray}
\sigma(E)=\pi\lambdabar(E)^2\frac{\omega\Gamma_\mathrm{i}(E)\Gamma_\mathrm{f}(E)}{(E-E_\mathrm{r})^2+[\Gamma_\mathrm{r}(E)/2]^2},\label{BW}
\end{eqnarray}
where $\lambdabar(E)$ is the de Broglie wave length, $\omega=(2J+1)/(2j_1+1)(2j_2+1)$ is the spin factor with $J$ the spin of the resonant state, and $j_1$ and $j_2$ the spins of the two nuclei in the entrance channel, and $\Gamma_\mathrm{i}(E)$ and $\Gamma_\mathrm{f}(E)$ are the partial widths for the entrance and exit channels, respectively. The energy dependence of the partial widths is given as $\Gamma_\mathrm{i,\;f}(E)=2P_\mathrm{i,\;f}(E)\gamma_\mathrm{i,\;f}^2$, where $P_\mathrm{i,\;f}(E)$ is the penetration factor and $\gamma_\mathrm{i,\;f}^2$ is the reduced width. The reaction rates are written as
\begin{equation}
  N_\mathrm{A}\langle\sigma v\rangle
  =\sqrt{\frac{8}{\pi\mu}}\frac{N_\mathrm{A}}{(k_\mathrm{B}T)^\frac{3}{2}}\int^\infty_0EdE\;\sigma(E)\mathrm{e}^{-E/k_\mathrm{B}T},\label{rate}
\end{equation}
where $\mu$ is the reduced mass of the two nuclei in the entrance channel. In the case of broad resonances, this integral should be performed numerically.

For the three reactions $^3$He$(d, p)^4$He, $^3$H$(d, n)^4$He, and $^7$Be$(n,p)^7$Li, the resonance energies are defined respectively as
\begin{eqnarray}
E^\mathrm{(d,p)}_\mathrm{r}&=&E_{^5\mathrm{Li}^\ast}-E_{^3\mathrm{He}}-E_\mathrm{d},\\
E^\mathrm{(d,n)}_\mathrm{r}&=&E_{^5\mathrm{He}^\ast}-E_\mathrm{t}-E_\mathrm{d},\\
E^\mathrm{(n,p)}_\mathrm{r}&=&E_{^8\mathrm{Be}^\ast}-E_{^7\mathrm{Be}}.
\end{eqnarray}
The shift of the resonance energy for the $^7$Be$(n,p)^7$Li reaction can then be written as
\begin{eqnarray}
\delta E^\mathrm{(n,p)}_\mathrm{r}&=&(K_{^8\mathrm{Be}^\ast}E_{^8\mathrm{Be}^\ast}-K_{^7\mathrm{Be}}E_{^7\mathrm{Be}})\frac{\delta m_q}{m_q}.
\end{eqnarray}
For other two reactions, the shifts are written in the same manner \cite{cheoun}.
The sensitivity coefficient $K_A$ for the excited states is not studied \cite{flambaum}, and we make assumptions of the following three cases separately for the three resonant reactions.

\paragraph{Case A} Variations in binding energies of excited states are the same as that of the ground state. In this case, the variation of the resonance energy is
\begin{eqnarray}
\delta E^\mathrm{(n,p)}_\mathrm{r}&=&(K_{^8\mathrm{Be}}E_{^8\mathrm{Be}}-K_{^7\mathrm{Be}}E_{^7\mathrm{Be}})\frac{\delta m_q}{m_q}.
\end{eqnarray}
\paragraph{Case B} The resonance height does not change in the reverse reaction  \cite{cheoun}. In this case, the variation of the resonance energy in the forward reaction is
\begin{eqnarray}
\delta E^\mathrm{(n,p)}_\mathrm{r}&=&(K_{^7\mathrm{Li}}E_{^7\mathrm{Li}}-K_{^7\mathrm{Be}}E_{^7\mathrm{Be}})\frac{\delta m_q}{m_q}.
\end{eqnarray}
\paragraph{Case C} The resonance energies do not change, i.e. $\delta E_\mathrm{r}=0$.

Our BBN calculation is based on Ref. \cite{kawano}. Reaction rates have been updated for $^4$He($\alpha$, $\gamma$)$^7$Be \cite{cyburt08} and $^2$H$(p, \gamma)^3$He ,$^2$H$(d, n)^3$He, and $^2$H$(d, p)^3$H \cite{coc15}. Other reaction rates are taken from the JINA REACLIB Database \cite{reaclib}.  We adopt the  neutron lifetime of 880.2 s \cite{pdg}. The baryon-to-photon ratio  $\eta=(6.108\pm0.060) \times10^{-10}$ is taken from Monte Carlo simulations \cite{cyburt} based on the Planck 2015 observational results \cite{planck}.

Table \ref{param} shows adopted resonance energies and widths for the $^7$Be$(n,p)^7$Li reaction at the present time \cite{descouvemont}.

\begin{table}
  \begin{tabular}{|ccccc|}
  \hline
$J$  & $E_\mathrm{r} $ [MeV]& $E_X $ [MeV]& $\Gamma_\mathrm{i}(E_\mathrm{r})$ [MeV]&$\Gamma_\mathrm{f} (E_\mathrm{r})$ [MeV]\\\hline
$2^-$ & 0.00267 & 18.91 & 0.225&1.41 \\
    $3^+$ & 0.330 & 19.07 & 0.0767 &0.088 \\
          &       & 19.24 &        &      \\
    $3^+$&2.66&21.5 & 0.490&0.610\\\hline
  \end{tabular}
\caption{The spins, resonance energies, corresponding excitation energies of $^8$Be$^\ast$, and partial widths for the resonances in the $^7$Be$(n,p)^7$Li reaction \cite{descouvemont}. \label{param}}
\end{table}

\begin{figure}
\includegraphics[width=8cm]{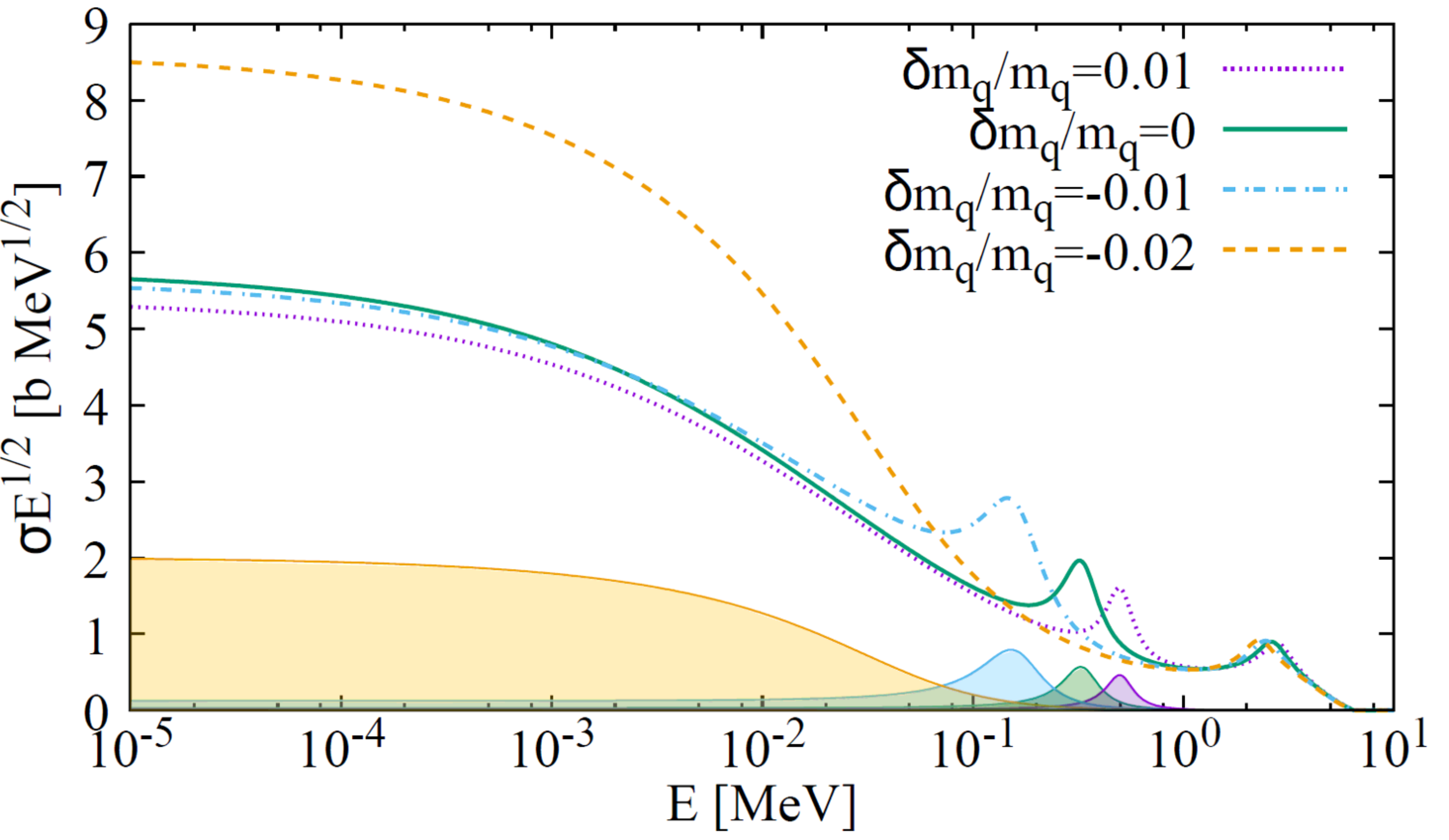}
\caption{\label{fig:sigma} The $^7$Be$(n,p)^7$Li cross sections multiplied by $E^{1/2}$ as a function of $E$ in Case A with quark mass variations as labeled. Shaded regions are contributions of the second resonant component scaled by 1/2.}
\end{figure}

Figure \ref{fig:sigma} shows calculated cross sections of the $^7$Be$(n,p)^7$Li reaction versus energy in Case A for $\delta m_q /m_q =$0.01 (dotted line), 0 (solid line), $-0.01$ (dashed-dotted line), and $-0.02$ (dashed line). The vertical axis $\sigma E^{1/2}$ is proportional to the reaction rate $\sigma v$ at the energy $E$. Shaded regions show partial cross sections by the second resonant component (Table \ref{param}) scaled by 1/2. In the energy range relevant to BBN, $E \lesssim {\mathcal O}(0.1)$ MeV, the first and second resonances predominantly contribute to the total cross section in SBBN.

In Case A, the resonance energy variation is $\delta E^\mathrm{(n,p)}_\mathrm{r}=17.2({\delta m_q}/{m_q})$ MeV. In the SBBN, the $^7$Be$(n,p)^7$Li cross section is predominantly contributed by a near-threshold resonance at $E_\mathrm{r}=2.67$ keV. However, the resonance energy of the next resonance at $E_\mathrm{r}=0.33$ MeV decreases to the kinetic energy at the $^7$Be synthesis, i.e. $E=3k_\mathrm{B}T/2\sim0.1$ MeV, if $\delta m_q/m_q$ is negative. As seen in Fig. \ref{fig:sigma}, all three resonances move to higher energies in the case of $\delta m_q /m_q =0.01$. The cross section is, therefore, smaller than in SBBN because of the hindered Boltzmann factor [Eq. (\ref{rate})] for $E \lesssim 0.3$ MeV above which the second resonance contributes to the total cross section. Cross sections at low energies are higher than in SBBN for $\delta m_q /m_q =-0.02$ because of significant contributions of the second excited state. We note that the lowest resonance at $E_\mathrm{r} =2.67$ keV in the present universe is always important for the total cross section in the all four cases of $\delta m_q/m_q$. Even when the resonant state becomes a subthreshold bound state, it contributes to the total cross section  because of its large width ($\Gamma =1.64$ MeV) and no hindrance of the neutron decay width $\Gamma_n$ from Coulomb potential.

\begin{figure}
\includegraphics[width=8cm]{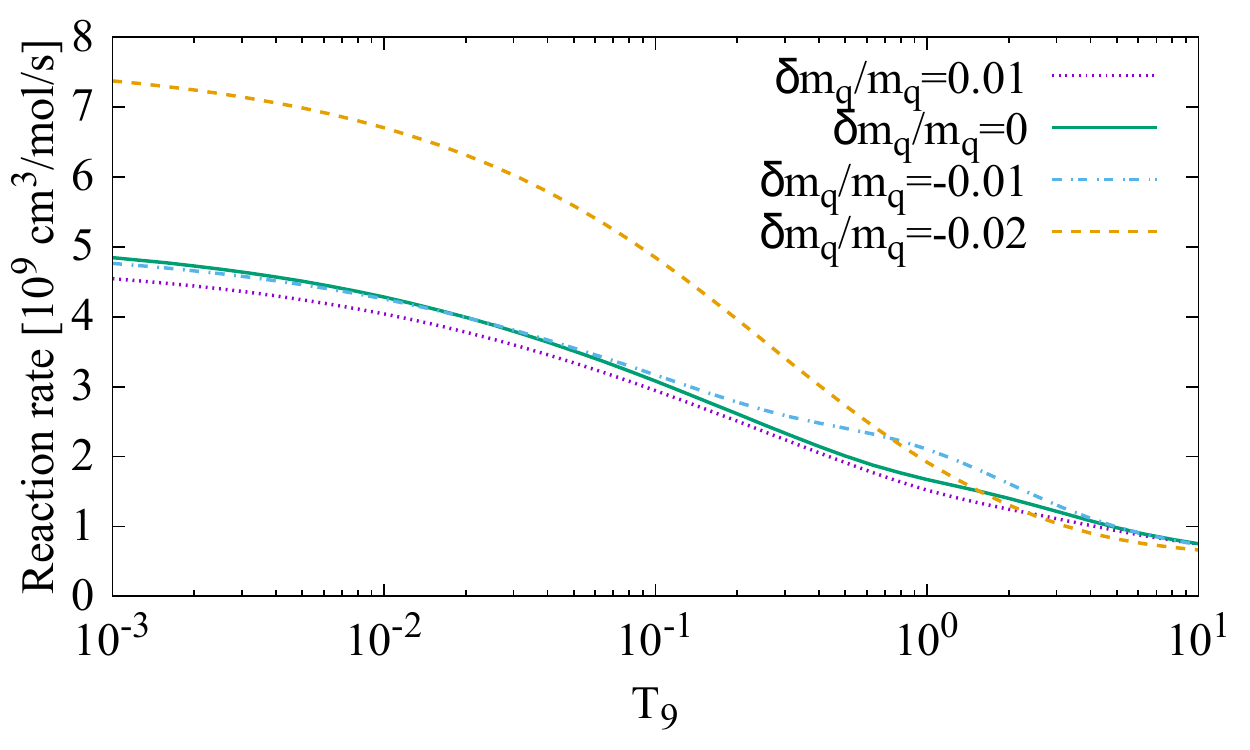}
\caption{\label{fig:rate} The $^7$Be$(n,p)^7$Li reaction rates versus $T_9 =T/(10^9~{\rm K})$ in Case A with the same quark mass variations as in Fig. \ref{fig:sigma}.}
\end{figure}

Figure \ref{fig:rate} shows the rate of the $^7$Be$(n,p)^7$Li reaction [Eq. (\ref{rate})] as a function of $T_9 =T/(10^9~{\rm K})$ for the same four cases as in Fig. \ref{fig:sigma}. In the temperature range most relevant to the $^7$Be destruction during BBN, i.e., $T_9 \lesssim 1$, the higher the quark mass is, the lower the reaction rate is. This results from the $\sigma v$ values at $E \lesssim 0.1$ MeV (see Fig. \ref{fig:sigma}). Thus, the reaction rates are increased for negative values of $\delta m_q /m_q$.

In Case B, the resonance energy variation is $\delta E^\mathrm{(n,p)}_\mathrm{r}=-0.172({\delta m_q}/{m_q})$ MeV. This shift is much smaller than in Case A, and the effect on the reaction rate is negligible.

In the $^3$He$(d,p)$ and $^3$H$(d,n)$ reactions, compound nuclei are expected to have similar energy levels from the mirror conjugate states. We then assume that the same Case is applied to these two resonant reactions. This assumption is not applicable to the $^7$Be$(n,p)^7$Li reaction. Because the resonant energy shift of the $^7$Be$(n, p)^7$Li reaction is negligible in Case B, Case B and C are almost the same for the  $^7$Be$(n,p)^7$Li reaction. Hence we consider only six cases (I to VI) shown in Table \ref{cases}.

\begin{table}
  \begin{tabular*}{80mm}{@{\extracolsep{\fill}}|c|ccc|}
  \hline
   \backslashbox{$^7$Be$(n,p)$}{$^3$He$(d,p)$, $^3$H$(d,n)$}   & A & B&C \\\hline
 A & I & III&V \\
    B/C & II & IV&VI \\\hline
  \end{tabular*}
\caption{The cases considered in this BBN calculation. A, B, and C are explained in text. The $^7$Be$(n,p)$ reaction is treated independently from the other reactions, because the structure of its compound nucleus is different from the others. \label{cases}}
\end{table}

\begin{figure}
\includegraphics[width=8cm]{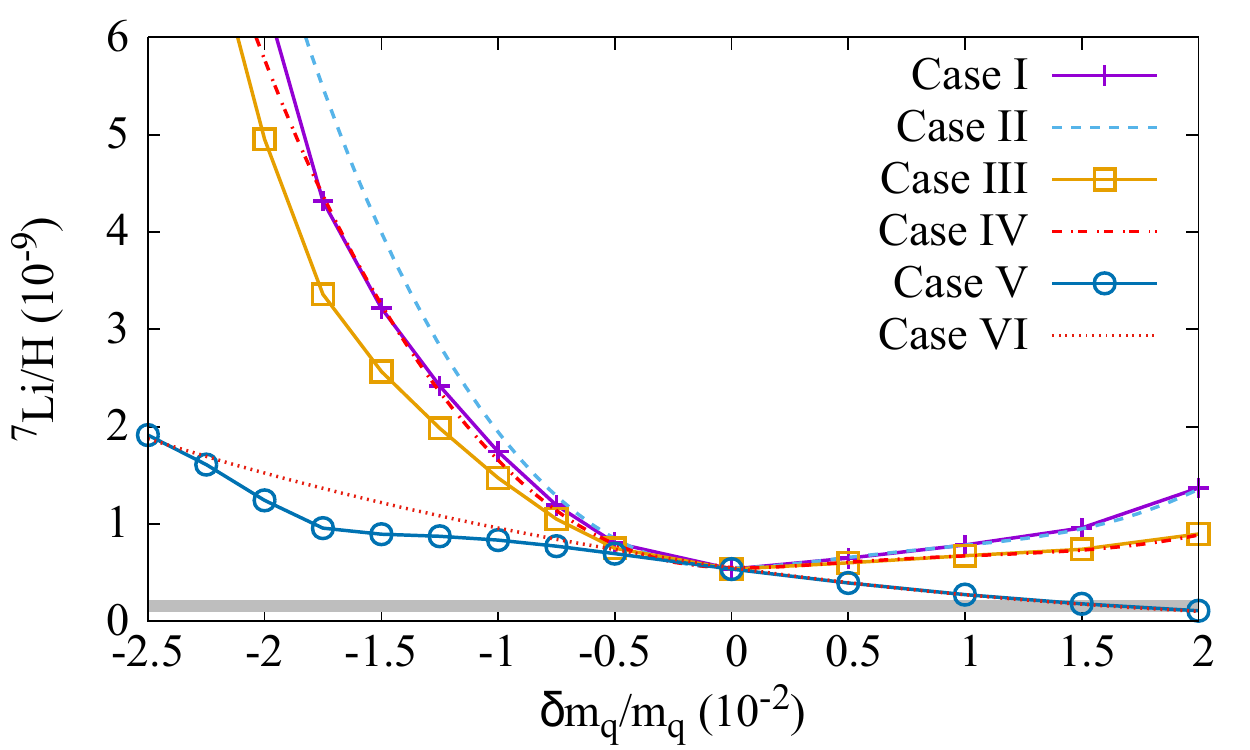}\\
\includegraphics[width=8cm]{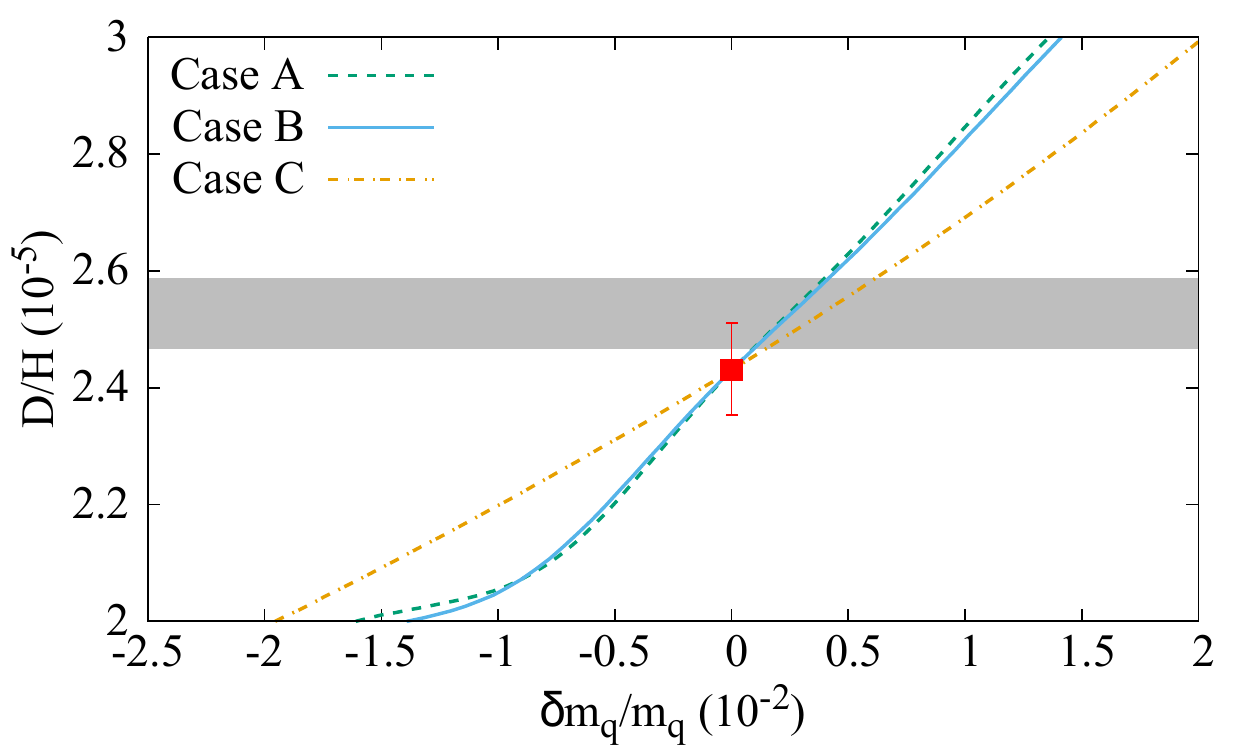}\\
\caption{\label{fig:bbn} (Upper panel) The $^7$Li abundance as a function of $\delta m_q/m_q$ in Cases I to VI (see Table \ref{cases}). The horizontal gray band shows the 2$\sigma$ range of abundances observed in metal-poor stars \cite{sbordone}. (Lower panel) The deuterium abundance as a function of $\delta m_q/m_q$ for Cases A, B and C for the $^3$He$(d, p)^4$He and $^3$H$(d, n)^4$He reactions. The gray band shows the  2$\sigma$ observational abundance \cite{cooke}. The solid square with an error bar shows the central value and the uncertainty in the SBBN.
}
\end{figure}

Figure \ref{fig:bbn} (upper panel) shows calculated Li abundance as a function of $\delta m_q/m_q$ for Cases I to VI. Cases II, IV, and VI are almost the same as cases investigated in Ref. \cite{cheoun} for which behaviors of nuclear abundances have been analyzed. $^7$Li abundances are determined by changes in reaction rates of (i) $^1$H($n$,$\gamma$)$^2$H, (ii) $^3$He($d$,$p$)$^4$He and $^3$H($d$,$n$)$^4$He, and (iii) $^7$Be($n$,$p$)$^7$Li. For larger $\delta m_q/m_q$ values, the $Q$-value of the reaction (i) becomes smaller. Due to a delayed deuteron production in BBN, the neutron abundance is higher. Since the neutron is the dominant $^7$Be destroyer via $^7$Be($n$,$p$)$^7$Li, the $^7$Be destruction rate is larger for larger $\delta m_q/m_q$. In addition, for larger $\delta m_q/m_q$ values, rates of reactions (ii) become smaller and the abundances of $^3$H and $^3$He become larger. As a result, abundances of $^7$Li and $^7$Be that are produced via $^3$H($\alpha$,$\gamma$) and $^3$He($\alpha$,$\gamma$) are larger. In previous studies, the rate of reaction (iii) has not been calculated accurately. However, it directly affects the $^7$Be and $^7$Li abundances.

The $^7$Li abundance decreases compared with Ref. \cite{cheoun} in Cases I, III, and V if $\delta m_q/m_q$ is negative. This is because the resonance of the $^7$Be$(n,p)^7$Li reaction at $E_\mathrm{r}=0.33$ MeV experiences an energy shift and the reaction rates at the BBN temperature are enhanced. The reaction $^7$Be$(n,p)^7$Li converts $^7$Be to $^7$Li, and $^7$Li nuclei are easily burnt via the reaction $^7$Li$(p,\alpha)^4$He. Thus the primordial $^7$Li abundance, which is the sum of abundances of $^7$Li and $^7$Be during the BBN, decreases. In Cases II, IV, and VI, the $^7$Li abundance agrees with the previous work \cite{cheoun}, because the  $^7$Be$(n,p)^7$Li resonant reaction rates are not significantly affected by the quark mass variation.

The gray band in the upper panel of Fig. \ref{fig:bbn} shows the Spite plateau \cite{spite} of $^7$Li abundance observed in metal-poor stars at $\log(\mathrm{Li}/\mathrm{H})+12=2.199\pm {0.086}$ \cite{sbordone}. Excepting the region of $\delta m_q/m_q \gtrsim 0.01$ in Cases V and VI, the observed level is lower than the calculated results of the BBN model.



Figure \ref{fig:bbn} (lower panel) shows the deuterium abundance as a function of $\delta m_q/m_q$. The gray band corresponds to the observational 2$\sigma$ range of primordial D abundance \cite{cooke}. It has been estimated from absorption lines in spectra of high-redshift quasars as $\mathrm{D/H}=(2.527\pm0.030)\times10^{-5}$.
The error bar around the solid square shows an uncertainty in the SBBN model originating from the $2\sigma$ uncertainty in $\eta$ \cite{cyburt}. The D abundance is larger for larger quark masses because of the delayed D production by the smaller $Q$ value of the $^1$H($n$,$\gamma$)$^2$H rate resulting in a larger freezeout abundance of D \cite{cheoun}.

\begin{figure}
\includegraphics[width=8cm]{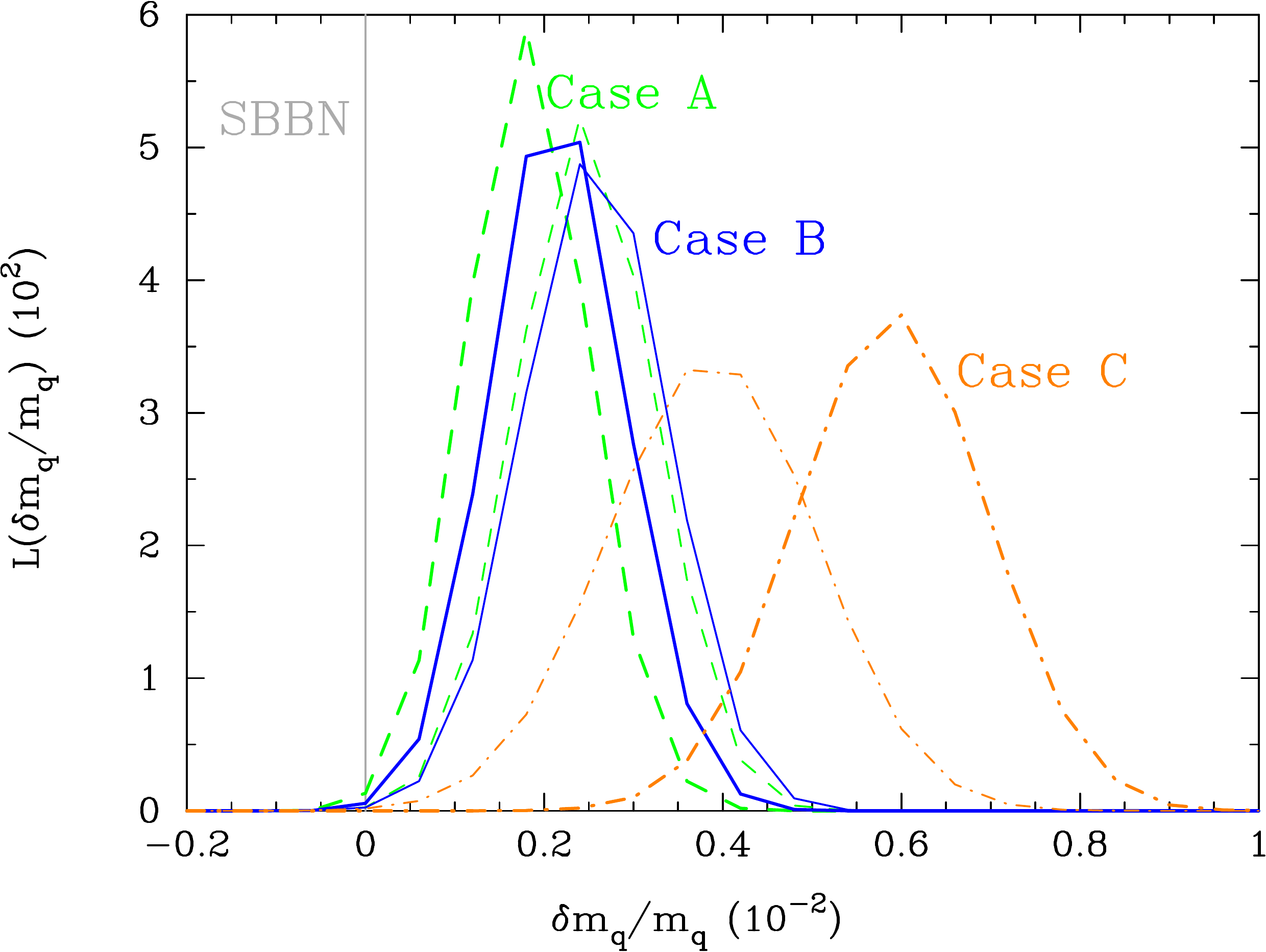}
\caption{\label{fig:likelihood} Likelihood functions of the quark mass variation $\delta m_q/m_q$ for Cases A to C of $^3$He$(d, p)^4$He and $^3$H$(d, n)^4$He. Thick and thin curves correspond to the functions derived with and without constraints from the observation of Li/H abundance, respectively. The vertical solid line on $\delta m_q/m_q=0$ corresponds to the SBBN model.}
\end{figure}

Figure \ref{fig:likelihood} shows the likelihood functions of the quark mass variation. Likelihood functions for respective nuclei $L_i$ ($i=$D, He, and Li) are assumed to be Gaussian, and the plotted ones are defined by products of $L_{\rm D} L_{\rm He} L_{\rm Li}$ (thick curves) and $L_{\rm D} L_{\rm He}$ (thin curves). We adopt the observational limit on the $^4$He abundance $Y_p=0.2449\pm0.0040$ \cite{aver}. When  the Li abundance is used as a constraint, best fit values of $\delta m_q/m_q$ move to lower $\delta m_q/m_q$ regions (Cases A and B) and a higher region (Case C) in directions of reducing the Li abundances. In Case C, the likelihood function is wider since the constraint from the D abundance is weaker (Fig. \ref{fig:bbn}).

Table \ref{limit} shows derived 95 \% C.L. for Case A to C. We find a solution to the Li problem with a finite value of $\delta m_q/m_q =($4--8$)\times 10^{-3}$ in Case C.
This result indicates that if the resonance energies of the reactions $^3$He$(d, p)^4$He and $^3$H$(d, n)^4$He are rather independent of the quark mass, the Li abundance becomes close to the observed value of metal-poor stars for $\delta m_q/m_q \lesssim 10^{-2}$.

\begin{table}
  \begin{tabular}{c|cccc}
  & D, He and Li & D and He \\\hline
Case A & 0.43--3.27 & 0.90--4.02 \\
Case B & 0.68--3.70 & 0.94--4.19 \\
Case C & 3.7--8.0 & 1.5--6.2 \\
  \end{tabular}
\caption{The 95 \% C.L. of $\delta m_q/m_q$ in units of $10^{-3}$ derived from observational constraints with (second column) and without (third column) Li abundances.\label{limit}}
\end{table}


In summary, the sensitivity of the $^7$Be$(n,p)^7$Li reaction rate to the change of a quark mass was investigated for the first time. We find that if the variation of the excitation energies of the compound nucleus $^8$Be$^\ast$ is the same as that of the ground state (i.e. Cases I, III, and V in Table \ref{cases}), the shift of the resonance at $E_{\rm r}=0.33$ MeV can decrease the $^7$Be abundance significantly for $\delta m_q/m_q \lesssim -5\times 10^{-3}$. 

The latest limit on the quark mass in the BBN epoch was derived for three models using updated observational constraints on primordial abundances. It is found that the Li abundance can be reduced toward the observed abundance level if the resonance energies of the reactions $^3$He$(d, p)^4$He and $^3$H$(d, n)^4$He does not depend on the quark mass.

This work is supported by NSFC Research Fund for International Young Scientists (11850410441).


\begin{thebibliography}{}
\bibitem{dirac} P. A. M. Dirac, Nature, 139, 323 (1937).
\bibitem{uzan} J. -P. Uzan, Living Rev. Relativity, 14, 2 (2011).
\bibitem{kolb} E. W. Kolb and M. S. Turner, The Early Universe (Basic Books, New York, 1994).
\bibitem{cyburt} R. H. Cyburt, B. D. Fields, K. A. Olive, and T. -H. Yeh, Rev. Mod. Phys., 88, 015004 (2016).
\bibitem{pitrou} C. Pitrou, A. Coc, J. -P. Uzan, and E. Vangioni, Phys. Rep., 754, 1 (2018).
\bibitem{Hinshaw:2012aka} 
  G.~Hinshaw {\it et al.} [WMAP Collaboration],
  Astrophys.\ J.\ Suppl.\  {\bf 208}, 19 (2013).
\bibitem{planck} P. A. R. Ade et al. (Planck Collaboration), Astron. Astrophys., 594, A13 (2016).
\bibitem{spite} F. Spite and M. Spite, Astron. Astrophys., 115, 357 (1982).
\bibitem{Ryan:1999vr}
  S.~G.~Ryan, T.~C.~Beers, K.~A.~Olive, B.~D.~Fields and J.~E.~Norris,
  Astrophys.\ J.\ {\bf 530}, L57 (2000).

\bibitem{Melendez:2004ni}
  J.~Melendez and I.~Ramirez,
  Astrophys.\ J.\  {\bf 615}, L33 (2004).

\bibitem{Asplund:2005yt}
  M.~Asplund, D.~L.~Lambert, P.~E.~Nissen, F.~Primas and V.~V.~Smith,
  Astrophys.\ J.\  {\bf 644}, 229 (2006).

\bibitem{Bonifacio:2006au} 
  P.~Bonifacio {\it et al.},
  Astron.\ Astrophys.\  {\bf 462}, 851 (2007).

\bibitem{shi2007}
  J.~R.~Shi, T.~Gehren, H.~W.~Zhang, J.~L.~Zeng, and G.~Zhao,
  Astron.\ Astrophys.\  {\bf 465}, 587 (2007).

\bibitem{Aoki:2009ce}
  W.~Aoki, P.~S.~Barklem, T.~C.~Beers, N.~Christlieb, S.~Inoue, A.~E.~G.~Perez, J.~E.~Norris and D.~Carollo,
  Astrophys.\ J.\  {\bf 698}, 1803 (2009).

\bibitem{Sbordone:2010zi}
  L.~Sbordone, P.~Bonifacio, E.~Caffau, H.~-G.~Ludwig, N.~T.~Behara, J.~I.~G.~Hernandez, M.~Steffen and R.~Cayrel {\it et al.},
  Astron.\ Astrophys.\  {\bf 522}, A26 (2010).
\bibitem{Cyburt:2009cf} 
  R.~H.~Cyburt and M.~Pospelov,
  Int.\ J.\ Mod.\ Phys.\ E {\bf 21}, 1250004 (2012).
  
\bibitem{Chakraborty:2010zj} 
  N.~Chakraborty, B.~D.~Fields and K.~A.~Olive,
  Phys.\ Rev.\ D {\bf 83}, 063006 (2011).
\bibitem{berengut} J. C. Berengut, V. V. Flambaum, and V, F, Dmitriev, Phys. Lett. B 683, 114 (2010).
\bibitem{cheoun} M. -K. Cheoun, T. Kajino, M. Kusakabe, and G. J. Mathews, Phys. Rev. D 84, 043001 (2011).
\bibitem{bedaque} P. F. Bedaque, T. Luu, and L. Platter, Phys. Rev. C 83, 045803 (2011).
\bibitem{descouvemont} P. Descouvemont, A. Adahchour, C. Angulo, A. Coc, and E. Vangioni-Flam, At. Data Nucl. Data Tables, 88, 203 (2004).
\bibitem{kawano} L. H. Kawano, FERMILAB-PUB-92 /04-A (1992).
\bibitem{exp1} R. L. Macklin, and J. H. Gibbons, Phys. Rev. 109, 105 (1958).
\bibitem{exp2} R. R. Borghers, and C. H. Poppe, Phys. Rev. 129, 2679 (1963).
\bibitem{exp3} C. H. Poppe, J. D. Anderson,  J. C. Davis, S. M. Grimes, and C. Wong, Phys. Rev. C 14, 438 (1976).
\bibitem{exp4} P. E. Koehler, C. D. Bowman, F. J. Steinkruger, D. C. Moody, G. M. Hale, J. W. Starner, S. A. Wender, R. C. Haight, P. W. Lisowski, and W. L. Talbert, Phys. Rev. C 37, 917 (1988).
\bibitem{exp5} L. Damone et al. [The n\textunderscore TOF Collaboration], Phys. Rev. Lett. 121, 042701 (2018).
\bibitem{exp6} S. Hayakawa et al., AIP Conf. Proc. 1947, 020011 (2018).
\bibitem{sbordone}  L. Sbordone et al., Astron. Astrophys., 522, A26 (2010).
\bibitem{cooke} R. J. Cooke, M. Pettini, and C. C. Steidel, Astrophys. J., 855, 102 (2018).
\bibitem{izotov} Y. I. Izotov, T. X. Thuan, and N. G. Guseva, Mon. Not. R. Astron. Soc., 445, 778 (2014).
\bibitem{aver} E. Aver, K .A. Olive, and E. D. Skillman, J. Cosmol. Astropart. Phys., 07, 011 (2015).
\bibitem{flambaum} V. V. Flambaum, and R. B. Wiringa, Phys. Rev. C 76, 054002 (2007).
\bibitem{iliadis} C. Iliadis, Nuclear Physics of Stars (Wiley-VCH, Weinheim, 2008).
\bibitem{cyburt08}R. H. Cyburt and B. Davids, Phys. Rev. C 78, 064614 (2008).
\bibitem{coc15} A. Coc et al., Phys. Rev. D 92, 123526 (2015).
\bibitem{reaclib} R. H. Cyburt et al., Astrophys. J. Suppl. Ser. 189, 240 (2010).
\bibitem{pdg} M. Tanabashi et al. (Particle Data Group), Phys. Rev. D 98, 030001 (2018). 
\end{thebibliography}
\end{document}



\title{Supplementary material}

\author{Kanji Mori}
\affiliation{%
Department of Astronomy, Graduate School of Science, The University of Tokyo, 7-3-1 Hongo, Bunkyo-ku,
Tokyo, 113-0033 Japan}%
\affiliation{%
National Astronomical Observatory of Japan, 2-21-1 Osawa, Mitaka, Tokyo,
181-8588 Japan}%

\author{Motohiko Kusakabe}
\affiliation{
School of Physics, Beihang University, 37 Xueyuan Road, Haidian-qu, Beijing
100083, China
}%

\date{\today}


\maketitle


\section{Observational constraints}\label{app1}
The constraints on the quark mass during the big bang nucleosynthesis (BBN) derived in this study are based on recent observational limit on the abundances of D, $^4$He and Li. Although observations of the primordial $^4$He abundance are taken into account in the likelihood analysis, they do not provide a strong constraint on the quark mass.
The primordial $^4$He abundance has been inferred from emission lines of metal-poor H\,{\sc ii} regions. It has been found that including not only visible lines but also infrared lines dissolves the degeneracy of physical parameters in spectral fittings, reducing uncertainties of the abundance \cite{izotov}. With this method, the $^4$He abundance is estimated as $Y_p=0.2449\pm0.0040$ \cite{aver}.

\begin{figure}
\includegraphics[width=8cm]{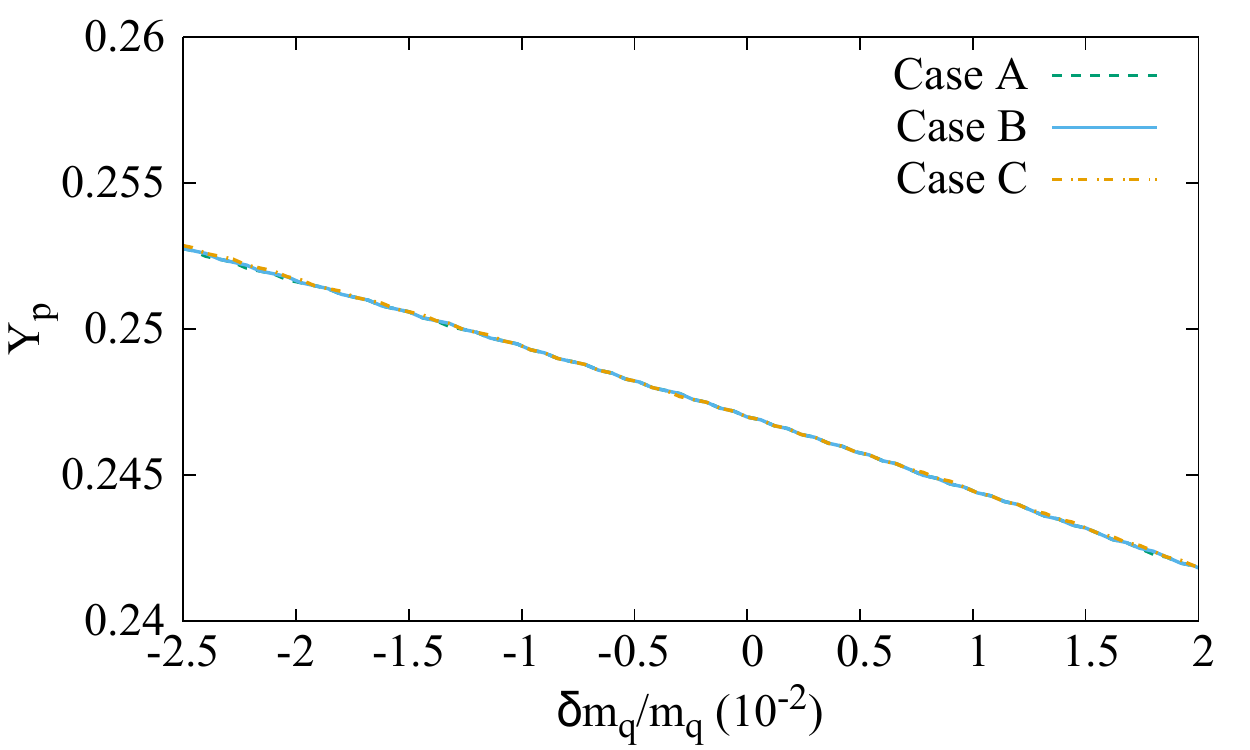}
\caption{\label{fig_app1} The $^4$He abundance as a function of $\delta m_q/m_q$ for Cases A, B and C for the $^3$He$(d, p)^4$He and $^3$H$(d, n)^4$He reactions. The whole region is allowed by the observational 2$\sigma$ limit on the $^4$He abundance \cite{aver}.}
\end{figure}

Figure \ref{fig_app1} shows calculated $^4$He abundance as a function of the quark mass variation $\delta m_q/m_q$. The $^4$He abundance is smaller for larger quark mass since the delayed D production leads to a smaller $n/p$ ratio at the D production or the $^4$He synthesis \cite{cheoun}. The whole region shown in this figure is consistent with the observed $^4$He abundance, and a stringent constraint on $\delta m_q/m_q$ is not derived from the $^4$He abundance.

\begin{figure}
\includegraphics[width=8cm]{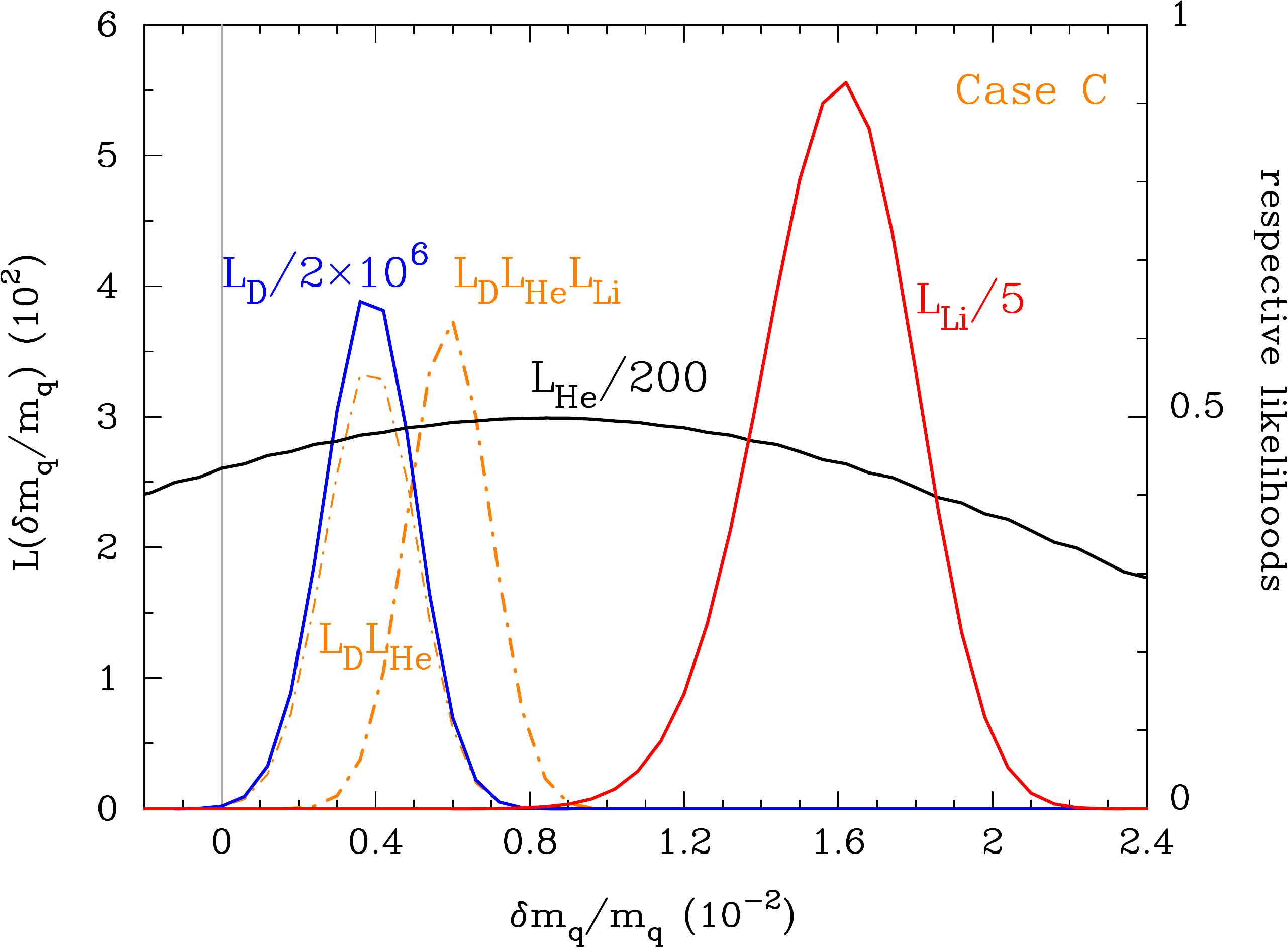}
\caption{\label{fig_app2} Likelihood functions of the quark mass variation $\delta m_q/m_q$ for Case C of the reactions $^3$He$(d, p)^4$He and $^3$H$(d, n)^4$He. Solid lines shows arbitrarily normalized likelihood functions for respective nuclei $L_i$. Thick and thin dashed-dotted lines are products of the functions $L_{\rm D}L_{\rm He}L_{\rm Li}$ and $L_{\rm D}L_{\rm He}$, respectively.}
\end{figure}

Figure \ref{fig_app2} shows the likelihood functions of the quark mass variation $\delta m_q/m_q$ for Case C, similar to figure 4 in the letter. Likelihood functions for respective nuclei $L_i$ ($i=$D, He, and Li) are assumed to be Gaussian, and thick and thin dashed-dotted lines correspond to $L_{\rm D} L_{\rm He} L_{\rm Li}$ and $L_{\rm D} L_{\rm He}$, respectively. Solid lines are likelihood functions for respective nuclei that are normalized arbitrarily as shown. It is observed that the likelihood $L_{\rm He}$ is high in the whole region, which indicates that the observational limit on the $^4$He abundance does not constrain this model of varying quark mass. Therefore, the best parameter region derived using the combined data on D and He abundances is predominantly determined by the D observational data. When the Li observation is additionally used as a constraint, the most likely parameter moves to a higher $\delta m_q/m_q$ region. The rapid drops of the likelihoods $L_{\rm D}$ and $L_{\rm Li}$ point to a finite change of the quark mass at $\delta m_q/m_q =($4--8$) \times 10^{-3}$.

\section{abundance evolution}

\begin{figure}[tbp]
\begin{center}
\includegraphics[width=8.0cm,clip]{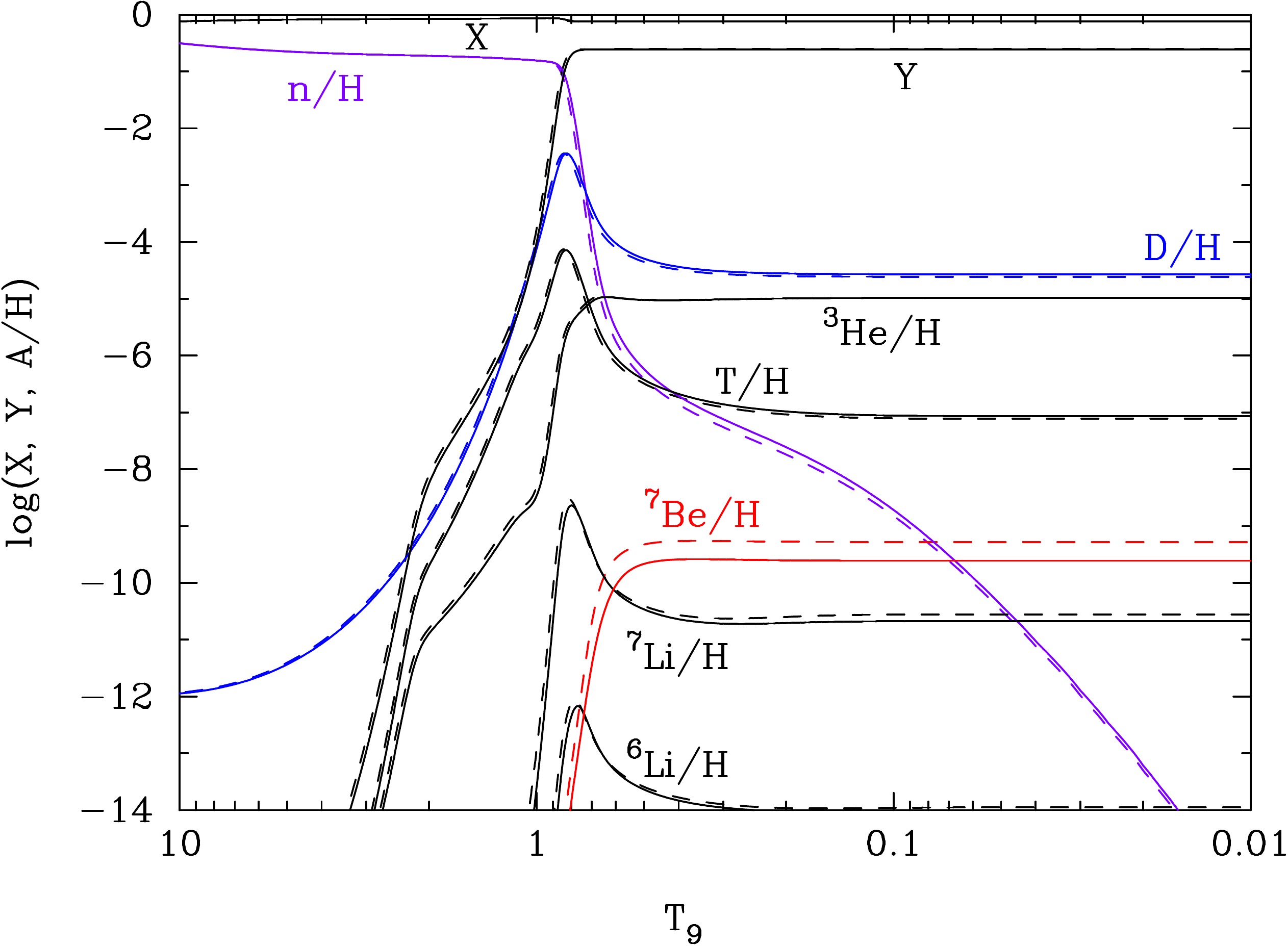}
\caption{Evolution of nuclear abundances as a function of $T_9=T/(10^9~{\rm K})$. $X$ and $Y$ are mass fractions of $^1$H and $^4$He in baryon in the universe, respectively. Other nuclear abundances are shown by the number ratios to $^1$H, i.e., $A$/H. The solid lines correspond to the heavy quark mass of $\delta m_q/m_q =10^{-2}$ for Case C, while the dashed lines correspond to $\delta m_q =0$, i.e., the SBBN model. \label{fig_app3}}
\end{center}
\end{figure}

Figure \ref{fig_app3} shows nuclear abundances as a function of $T_9=T/(10^9~{\rm K})$. $X$ and $Y$ are mass fractions of $^1$H and $^4$He in baryon in the universe, respectively. Other nuclear abundances are shown by the number ratios to $^1$H, i.e., $A$/H. The solid lines correspond to the heavy quark mass of $\delta m_q/m_q =10^{-2}$ for Case C, which is the most interesting parameter region with a reduced Li abundance. The dashed lines correspond to $\delta m_q =0$. In the heavy mass case, the neutron and D abundances are higher and the $^7$Be abundance is lower than in the standard BBN (SBBN). 


As described in the letter, the important reactions for the $^7$Be abundance during BBN or the primordial Li abundance are (i) $^1$H($n$,$\gamma$)$^2$H, (ii) $^3$He($d$,$p$)$^4$He and $^3$H($d$,$n$)$^4$He, and (iii) $^7$Be($n$,$p$)$^7$Li. 
If $\delta m_q/m_q$ is large, the $Q$-value of the reaction (i) is smaller than in SBBN. As a result, the D production is slightly delayed and the neutron abundance is higher. This high abundance of neutrons destroy more $^7$Be nuclei via the $^7$Be($n$,$p$)$^7$Li reaction. Although the reactions (ii) are also important, their resonance energies are unchanged from the SBBN values in Case C. Therefore, there is no significant change in the abundances of either $^3$H or $^3$He. The rate of reaction (iii) is slightly smaller than in SBBN (figure 2 in the letter), which reduces the efficiency of $^7$Be destruction. However, because of the increased $n$ abundance, the higher $^7$Be destruction rate effectively delays the accumulation of $^7$Be. Therefore, the net effect in Case C for $\delta m_q/m_q =10^{-2}$ is a significant reduction of the final $^7$Be abundance. We note that a slight increase in the D abundance seen in this figure is constrained from the D observation (see Fig. \ref{fig_app2}).
